\documentclass[11pt,a4paper]{article}
\usepackage{amsmath,amssymb,siunitx,geometry}
\usepackage{booktabs}
\usepackage{hyperref}
\usepackage{physics}
\usepackage{graphicx}
\usepackage{caption}
\usepackage{siunitx}
\title{A general thermodynamic model for the steady-state temperature of a photovoltaic module on the Moon}
\author{Mykhailo Koltsov\thanks{Laboratory for Thin Film Energy Materials: Department of Materials and Environmental Technology, Tallinn University of Technology, Tallinn, Estonia. \href{mailto:mykhailo.koltsov@taltech.ee}{mykhailo.koltsov@taltech.ee}}
\and Zacharie Jehl Li-Kao\thanks{MNT Photovoltaic Group, Department of Electronic Engineering, Universitat Politècnica de Catalunya, Barcelona, Spain. Corresponding author: \href{mailto:zacharie.jehl@upc.edu}{zacharie.jehl@upc.edu}}}
\date{}

\begin{document}
\maketitle

\begin{abstract}
Photovoltaic conversion is highly dependent on the converter's temperature. In the absence of atmosphere, it can be rigorously determined using a thermal balance leading to a compact expression. The derivation starts from a fully spectral, two-sided radiative balance and proceeds to a practical gray-band (integrated) expression that can be used to estimate Tc from measurable quantities: plane-of-array irradiance, spectral absorptance (or a broadband absorptance), thermal emissivities, view factors to the regolith and to space, module electrical efficiency, and conduction to the mount. The model can be extended to other airless celestial bodies and to deep space.

\end{abstract}

\section{Introduction}
Several models for estimating the cell temperature of solar photovoltaic (PV) converters have been developed for terrestrial application \cite{de_soto_improvement_2006,faiman_assessing_2008}. Due to the complexity of the problem and the large number of interplaying factors, these models rely on empirical coefficients. These coefficients capture effects such as convective cooling, radiative losses, and wind influence, allowing the models to predict module temperature without solving the full heat transfer problem from first principles.
On a celestial body like the Moon, the absence of an atmosphere significantly simplifies the thermal problem and convective heat transfer can be ignored. Under fixed solar illumination conditions, the irradiance on the plane of the array can be approximated as constant as well. Current approaches to lunar photovoltaic system modelling address complementary but distinct aspects of the design challenge, each with their own assumptions and limitations. Mission-specific engineering optimization studies \cite{breeding_optimization_2018} recognize thermal challenges but rely on simplified radiative models that lack systematic treatment of regolith thermal emission, reflected solar radiation, and rigorous view factor calculations. \cite{silva-rodriguez_solar_2024} incorporate temperature-dependent efficiency by relating PV cell temperature to lunar ground temperature, which implicitly includes regolith heating effects. However, their approach does not explicitly resolve radiative exchange with the regolith or space, nor does it account for geometry- and wavelength-dependent factors that can significantly influence module thermal balance. While these approaches offer useful frameworks for solar resource assessment, geometric optimisation, and mission-specific design considerations, they may be limited in their ability to accurately predict operating temperatures or performance under actual lunar thermal conditions.

Our comprehensive thermal model aims to address these fundamental gaps through a systematic derivation starting from fully spectral energy balance equations, then simplified to practical gray-band expressions while maintaining physical rigour. The method explicitly treats two-sided radiative heat transfer with distinct front and back face optical properties ($\alpha_f$, $\alpha_b$, $\varepsilon_f$, $\varepsilon_b$), incorporates geometric view factors ($F_r$, $F_g$) for both reflected solar radiation and regolith thermal IR emission, includes conduction to mounting structures, and provides a clear framework for extending to spectral accuracy when needed. The resulting compact formula can help with rapid temperature prediction from measurable quantities (such as plane-of-array irradiance, optical properties, view factors, ambient conditions) while also capturing the dominant physical processes that other models may overlook. However, it should be stressed that our approach assumes steady-state conditions, and it employs single-node isothermal approximations. Complete lunar PV system design requires integration of thermal physics modelling with geometric optimisation and other solar resource assessment methodologies.

\section{Assumptions of the model}
This model aims for a steady-state (time derivatives set to zero) for a PV module that exchanges energy only by radiation and conduction (no convection in vacuum). The primary assumptions are:
\begin{itemize}
\item Vacuum environment: no convective heat transfer.
\item Module approximated as a single isothermal body at temperature $T_c$ (first-order approximation). A two-layer model (glass + cell) is possible but not developed here.
\item The Sun provides direct beam irradiance (AM0) with plane-of-array irradiance $G_{\text{POA}}$ (W m$^{-2}$). There is negligible atmospheric scattering.
\item Lunar regolith emits thermally as a body at temperature $T_{\text{reg}}$ (K) and reflects sunlight with broadband albedo $\rho$ (we assume a Lambertian reflection assumed for geometric factors).
\item Spectral properties may be reduced to integrated (broadband) coefficients when appropriate: solar absorptance $\alpha$, thermal emissivity $\varepsilon$, etc. Spectral integrals and definitions are provided below.
\item Electrical power extracted by the PV module is $P_{\text{elec}} = \eta G_{\text{POA}}$ per unit module area, where $\eta$ is module efficiency (which in the general case depends on $T_c$).
\item Conduction into the mounting structure may be represented by a thermal resistance per unit area $R_{\text{th}}$.
\end{itemize}

\section{Notation and definitions}
We list here the symbols used through this paper. All fluxes are expressed per unit geometric module area (W m$^{-2}$).

\begin{table}[h]
\centering
\begin{tabular}{p{3cm}p{10cm}}
\toprule
\textbf{Symbol} & \textbf{Definition} \\
\midrule
$G$ & solar irradiance at 1 AU normal to the Sun (W m$^{-2}$), typically $\approx 1361$ W m$^{-2}$ \\
$G_{\text{POA}}$ & plane-of-array direct irradiance incident on module face (W m$^{-2}$), $G_{\text{POA}}=G\cos\theta$ for a flat plate at incidence angle $\theta$ \\
$\alpha_{f}, \alpha_{b}$ & broadband solar absorptance of front and back faces (unitless; fraction of incident solar power absorbed) \\
$\varepsilon_{f}, \varepsilon_{b}$ & broadband thermal emissivity of front and back faces in the mid/far IR (unitless) \\
$\varepsilon_{\text{reg}}$ & regolith broadband thermal emissivity in the mid/far IR (unitless) \\
$\eta$ & electrical conversion efficiency of the module referenced to incident POA (unitless); $P_{\text{elec}}=\eta G_{\text{POA}}$ \\
$\rho$ & regolith broadband albedo in the solar band (unitless) \\
$F_{r,f}, F_{r,b}$ & geometric coefficients for regolith-reflected solar irradiance incident on the front/back faces (unitless, 0--1) \\
$F_{g,f}, F_{g,b}$ & view factors from the front/back faces to the regolith for thermal IR exchange (unitless, 0--1) \\
$T_{\text{reg}}$ & regolith surface temperature (K) \\
$T_c$ & module temperature (K) (unknown to solve for) \\
$\sigma$ & Stefan--Boltzmann constant, $\sigma = 5.670374419\times10^{-8}$ W m$^{-2}$ K$^{-4}$ \\
$q_{\text{cond}}$ & net conductive flux from module to mount per unit area (W m$^{-2}$). If modeled via thermal resistance per unit area $R_{\text{th}}$, then $q_{\text{cond}}=(T_c - T_{\text{mount}})/R_{\text{th}}$ \\
\bottomrule
\end{tabular}
\caption{Notation and symbols used throughout the document. For each face, $F_{g,i}$ is the fraction of the hemisphere viewed by regolith; the complementary fraction $1-F_{g,i}$ corresponds to the view factor to space. The coefficients $F_{r,i}$ quantify the fraction of regolith-reflected solar irradiance that reaches face $i$ and depend on geometry (tilt, mounting height, surroundings).}
\end{table}

For a simple first approximation, that is an infinite, flat regolith surface with Lambertian reflection and a module tilted by angle $\beta$ from horizontal, the ground-view factor for the front face may be taken as
\[
F_{g,\text{front}} \approx \frac{1-\cos\beta}{2}
\]
with the complementary sky-view factor $F_{\text{space,front}}=1-F_{g,\text{front}}$. This closed form fails at low mounting heights or in the presence of nearby structures; in those cases compute view factors by direct integration or a radiosity method.

\section{Spectral start: full spectral energy balance}
A general starting point is a spectral radiative balance per unit area for the entire module on both faces. We define $\lambda$ as being the wavelength, $I_{\odot}(\lambda)$ the AM0 spectral irradiance, and $B(\lambda,T)$ the Planck spectral exitance per unit wavelength (W\,m$^{-2}$\,$\mu$m$^{-1}$).

The spectral absorptance of the module is $A(\lambda)$ and its spectral emissivity is $\varepsilon(\lambda)$; according to Kirchhoff's law these are equal for an opaque surface at the same wavelength and direction.

The steady-state spectral balance integrated over all $\lambda$ is:
\begin{equation}\label{eq:spectral_balance}
\int_{0}^{\infty} A(\lambda) I_{\odot,\text{POA}}(\lambda) \,\mathrm{d}\lambda
+ \int_{0}^{\infty} A(\lambda) E_{\text{reg}}(\lambda) \,\mathrm{d}\lambda
- \int_{0}^{\infty} \varepsilon(\lambda) B(\lambda,T_c) \,\mathrm{d}\lambda
- P_{\text{elec}}
- q_{\text{cond}}
= 0
\end{equation}
where $I_{\odot,\text{POA}}(\lambda)$ is the spectral irradiance incident on the illuminated face (includes incidence cosine correction and projected area effects), and $E_{\text{reg}}(\lambda)$ is the spectral irradiance from the regolith on the module. It should be specified that $E_{\text{reg}}(\lambda)$ includes thermal emission from the regolith and reflected solar. More explicitly, for regolith thermal emission we can also write
\begin{equation}
E_{\text{reg,th}}(\lambda)
= \varepsilon_{\text{reg}}(\lambda) B(\lambda,T_{\text{reg}}) F_{\text{view}}(\text{reg} \rightarrow \text{module})
\end{equation}
and for reflected solar from regolith
\begin{equation}
E_{\text{reg,ref}}(\lambda)
= \rho(\lambda) I_{\odot}(\lambda) F_{r}(\text{geometry})
\end{equation}

Equation~\eqref{eq:spectral_balance} is the most general steady-state statement: all absorbed spectral radiance from solar emission and the regolith is balanced by the module emission, electrical extraction, and conduction.

\section{Gray-band (integrated) simplification}
For practical use we can collapse spectral integrals into a few broadband coefficients. We then define the following integrated quantities (all normalized per unit module area):
\begin{align}
G_{\text{POA}} &\equiv \int_{\text{solar band}} I_{\odot,\text{POA}}(\lambda) \,\mathrm{d}\lambda \quad\text{(W m}^{-2}\text{)} \\
\alpha_{i} &\equiv \frac{1}{G_{\text{POA}}}\int_{\text{solar band}} A_{i}(\lambda) I_{\odot,\text{POA}}(\lambda) \,\mathrm{d}\lambda \quad (i=\text{f, b}) \\
\varepsilon_{i} &\equiv \frac{1}{\sigma T_c^{4}}\int_{\text{IR band}} \varepsilon_{i}(\lambda) B(\lambda,T_c) \,\mathrm{d}\lambda \quad (i=\text{f, b})
\end{align}
Here $i=\text{f}$ or $\text{b}$ is for front/back faces. The solar-band absorptance $\alpha_i$ is weighted by the AM0 spectral shape; the thermal emissivity $\varepsilon_i$ is defined so that $\varepsilon_i\sigma T_c^4$ equals the integrated emitted IR flux from face $i$ (but strictly speaking, $\varepsilon_i$ depends weakly on $T_c$ because the Planck distribution shifts with temperature).

Using these definitions, the absorbed solar power per unit module area can now be written as follows
\begin{equation}
q_{\text{sol,abs}} = \alpha_f G_{\text{POA}}
+ \alpha_f \rho G\, F_{r,f}
+ \alpha_b \rho G\, F_{r,b}
\end{equation}
Where the first term is direct POA on the front and the other two are reflected solar captured by each face according to geometry.

The absorbed thermal power from the regolith is
\begin{equation}
q_{\text{reg,abs}} = \varepsilon_{\text{reg}}\bigl(\varepsilon_f \sigma F_{g,f} T_{\text{reg}}^{4} + \varepsilon_b \sigma F_{g,b} T_{\text{reg}}^{4}\bigr)
\end{equation}
The emitted thermal power from the module (both faces) is
\begin{equation}
q_{\text{emit}} = \varepsilon_f \sigma T_c^{4} + \varepsilon_b \sigma T_c^{4} = (\varepsilon_f+\varepsilon_b) \sigma T_c^{4}
\end{equation}

Electrical extraction per unit module area (basically the power removed from the thermal balance) has already been defined as
\begin{equation}
P_{\text{elec}} = \eta G_{\text{POA}}
\end{equation}

Conduction to the mounting structure is modelled as
\begin{equation}
q_{\text{cond}} = \frac{T_c - T_{\text{mount}}}{R_{\text{th}}}
\end{equation}
where $R_{\text{th}}$ (K m$^{2}$ W$^{-1}$) is the thermal resistance per unit area of the mount path and $T_{\text{mount}}$ is the temperature of the heat sink (K). If the structure is large and acts as an infinite sink at $T_{\text{mount}}$, this linear model is appropriate. If the module is effectively floating, $R_{\text{th}}\to\infty$ and $q_{\text{cond}}\to 0$.
Here $q_{\text{cond}}$ is defined positive when heat flows out of the module into the mount, so it appears as a loss term in the balance.

\section{Two-sided integrated steady-state balance}
We can now combine the terms into the steady-state balance:
\begin{multline}\label{eq:balance_integrated}
\underbrace{\alpha_f G_{\text{POA}} + \alpha_f \rho G F_{r,f} + \alpha_b \rho G F_{r,b}}_{\text{absorbed solar}}
+ \underbrace{\varepsilon_{\text{reg}}\big(\varepsilon_f\sigma F_{g,f} T_{\text{reg}}^{4} + \varepsilon_b\sigma F_{g,b} T_{\text{reg}}^{4}\big)}_{\text{regolith thermal in}} \\
= \underbrace{(\varepsilon_f+\varepsilon_b)\sigma T_c^{4}}_{\text{module emission}}
+ \underbrace{\eta G_{\text{POA}}}_{\text{electrical extraction}}
+ \underbrace{q_{\text{cond}}}_{\text{conduction}}
\end{multline}

And we then rearrange to isolate the $T_c^{4}$ term:
\begin{equation}\label{eq:Tc4_general}
T_c^{4} = \frac{\alpha_f G_{\text{POA}} + \alpha_f \rho G F_{r,f} + \alpha_b \rho G F_{r,b} + \varepsilon_{\text{reg}}\big(\varepsilon_f\sigma F_{g,f} T_{\text{reg}}^{4} + \varepsilon_b\sigma F_{g,b} T_{\text{reg}}^{4}\big) - \eta G_{\text{POA}} - q_{\text{cond}}}{(\varepsilon_f+\varepsilon_b)\sigma}
\end{equation}

Thus a compact expression for the module temperature $T_c$ is
\begin{equation}\label{eq:Tc_summary}
\boxed{
T_c = \left[ \frac{(\alpha_f - \eta) G_{\text{POA}} + \alpha_f \rho G F_{r,f} + \alpha_b \rho G F_{r,b} + \varepsilon_{\text{reg}}\big(\varepsilon_f\sigma F_{g,f} T_{\text{reg}}^{4} + \varepsilon_b\sigma F_{g,b} T_{\text{reg}}^{4}\big) - q_{\text{cond}}}{(\varepsilon_f+\varepsilon_b)\sigma} \right]^{1/4}}
\end{equation}

Equation~\eqref{eq:Tc_summary} represents the practical gray-body formulation, and it is the main result of this work. A Python model is provided as supplementary material to illustrate the model and to visualise the dependence of $T_c$ on the variables from Equation~\eqref{eq:Tc_summary}.
This equation is explicit if the electrical efficiency $\eta$ dependence on the temperature is known, and if the conductive flux $q_{\text{cond}}$ is a known constant. At the first order, the dependence of $\eta$ on the temperature can be estimated using moelling tools such as SCAPS or SOLEY, but experimental data should probably be favoured in this case. If in a situation where $\eta$ depends in an unknown manner on $T_c$, so Equation~\eqref{eq:Tc_summary} becomes implicit and must be solved iteratively. In the following, we explain how to interpret each term and how to extend the model for transient and spectral accuracy.

\subsection*{Interpretation of terms}
\begin{itemize}
  \item $(\alpha_f-\eta) G_{\text{POA}}$: net solar heating on the front face. $\alpha_f G_{\text{POA}}$ is the solar power absorbed by the front; $\eta G_{\text{POA}}$ is the portion of incident solar power converted to electricity. The difference is the solar power that thermalises in the module.
  \item $\alpha_f \rho G F_{r,f}$ and $\alpha_b \rho G F_{r,b}$: heating produced by sunlight reflected from the regolith (albedo) and captured by front and back according to geometry (the $F_r$ factors).
  \item $\varepsilon_{\text{reg}}\big(\varepsilon_f\sigma F_{g,f}T_{\text{reg}}^{4} + \varepsilon_b\sigma F_{g,b}T_{\text{reg}}^{4}\big)$: thermal flux from the warm regolith received by the faces. If $\varepsilon_{\text{reg}}\approx 1$ this prefactor may be dropped. Because $\sigma T^4$ grows rapidly with $T$, a regolith at $\sim$350--400 K can supply thermal fluxes comparable to solar irradiance if the view factors are large.
  \item $(\varepsilon_f+\varepsilon_b)\sigma$: combined radiative loss capacity of both faces. Larger emissivity increases radiative cooling and lowers steady $T_c$ for fixed inputs.
  \item $q_{\text{cond}}$: conductive sink to the mount (may be zero if module is thermally isolated). 
\end{itemize}

\subsection*{Important of the spectral accuracy}
Equation~\eqref{eq:Tc_summary} uses broadband coefficients. But in reality:
\begin{itemize}
  \item Regolith emission is centered in the mid/far-IR (Wien peak $\lambda_{\mathrm{max}}\approx 2898/T$ in $\mu$m; for $T\sim 390$ K, $\lambda_{\mathrm{max}}\approx 7.4\ \mu$m).
  \item Typical PV bandgaps (Si $\sim 1.12$ eV) cut off at $\lambda \approx 1.1\ \mu$m, so most regolith photons cannot generate carriers; they are absorbed in glass/encapsulant and heat the module.
  \item Glass and encapsulant optical properties vary strongly with wavelength: high transmittance in solar band, strong absorption in mid-IR. Kirchhoff's law implies those layers also emit strongly in IR.
\end{itemize}
Therefore, if we need to quantify how much of regolith IR reaches the semiconductor junction vs being absorbed by the glass, we should implement the spectral balance
\begin{equation}
\int A(\lambda) I_{\odot,\text{POA}}(\lambda)\,d\lambda + \int A(\lambda)E_{\text{reg}}(\lambda)\,d\lambda
= \int \varepsilon(\lambda) B(\lambda,T_c)\,d\lambda + P_{\text{elec}} + q_{\text{cond}}
\end{equation}
numerically integrating with measured or representative $A(\lambda)$ and $\varepsilon(\lambda)$. This is of course beyond the scope of this work.

\subsection*{Transient thermal model including thermal inertia}

To extend the model to time-dependent conditions, we introduce an areal thermal capacitance $C = m c$ (in J\,m$^{-2}$\,K$^{-1}$), representing the module’s ability to store heat, where $m$ is the module mass per unit area and $c$ is the specific heat capacity of the module materials. The transient energy balance for the module is then:

\begin{equation}\label{eq:transient_balance}
\begin{split}
C \frac{\mathrm{d}T_c}{\mathrm{d}t} 
&= \underbrace{\alpha_f G_{\text{POA}} + \alpha_f \rho G F_{r,f} + \alpha_b \rho G F_{r,b}}_{\text{absorbed solar flux}}
+ \underbrace{\varepsilon_{\text{reg}}\big(\varepsilon_f\sigma F_{g,f} T_{\text{reg}}^{4} + \varepsilon_b\sigma F_{g,b} T_{\text{reg}}^{4}\big)}_{\text{regolith thermal flux}} \\
&\quad - \underbrace{(\varepsilon_f+\varepsilon_b)\sigma T_c^4}_{\text{radiative losses}} 
- \underbrace{\eta(T_c) G_{\text{POA}}}_{\text{electrical extraction}} 
- \underbrace{\frac{T_c-T_{\text{mount}}}{R_{\text{th}}}}_{\text{conductive losses}}
\end{split}
\end{equation}

Equation~\eqref{eq:transient_balance} can be integrated numerically using solvers to predict the time evolution of $T_c(t)$. For slowly varying conditions (the lunar day-night cycle is considered slow in this case), the characteristic thermal response time of the module can be estimated as:

\begin{equation}
\tau \approx \frac{C}{4(\varepsilon_f+\varepsilon_b)\sigma T_c^3 + 1/R_\text{th}}
\end{equation}

which represents the first-order time constant of the system around a steady-state temperature $T_c$.  This formulation naturally reduces to the steady-state expression in Equation~\eqref{eq:Tc_summary} when $\mathrm{d}T_c/\mathrm{d}t = 0$, and provides a framework to include transient thermal effects such as thermal inertia, diurnal variation, or variable irradiance.

\subsection*{Simple numerical application}
We can equation~\eqref{eq:Tc_summary} with the following representative values:
\begin{align}
&G = 1361\ \text{W m}^{-2}, \quad G_{\text{POA}}=G\ (\cos\theta=1), \nonumber \\
&\alpha_f = 0.90, \quad \alpha_b = 0.05, \quad \varepsilon_f = \varepsilon_b = 0.90, \nonumber \\
&\eta = 0.20, \quad \rho = 0.12, \quad F_{r,f}=0.10, \quad F_{r,b}=0, \nonumber \\
&F_{g,f}=F_{g,b}=0.10, \quad T_{\text{reg}}=390\ \text{K}, \quad q_{\text{cond}}=0. \nonumber
\end{align}
Compute contributions:
\begin{align}
(\alpha_f-\eta)G_{\text{POA}} &=(0.90-0.20)\times1361 \approx 952.7\ \text{W m}^{-2}, \nonumber \\
\alpha_f\rho G F_{r,f} &\approx 0.90\times0.12\times1361\times0.10 \approx 14.7\ \text{W m}^{-2}, \nonumber \\
\sigma T_{\text{reg}}^{4} &\approx 5.67\times10^{-8}\times390^4 \approx 1312\ \text{W m}^{-2}, \nonumber \\
\varepsilon_f\sigma F_{g,f}T_{\text{reg}}^{4}+\varepsilon_b\sigma F_{g,b}T_{\text{reg}}^{4}
&\approx 0.9\times1312\times0.1 + 0.9\times1312\times0.1 \approx 236\ \text{W m}^{-2}. \nonumber
\end{align}
Sum numerator $\approx 952.7+14.7+236 \approx 1203\ \text{W m}^{-2}$. Denominator $(\varepsilon_f+\varepsilon_b)\sigma \approx 1.02\times10^{-7}\ \text{W m}^{-2}\text{K}^{-4}$. Thus
\begin{equation}
T_c \approx \left(\frac{1203}{1.02\times10^{-7}}\right)^{1/4} \approx 329\ \text{K}\ (\approx 56^\circ\text{C})
\end{equation}
This quick calculation demonstrates regolith IR adds a non-negligible term and that the gray-body formula yields plausible temperatures for a PV module on the moon.

\subsection*{Limitations of this approach}
The one-node isothermal model neglects in-plane temperature gradients, which can be significant for large panels or in situations with large edge losses. In such cases, a two-dimensional conduction model or multiple thermal nodes could be considered. Accurate spectral modelling is necessary to determine where infrared energy is absorbed, whether in the glass or directly in the cell, and to estimate how much contributes to heating the junction versus heating the encapsulant. And of course, dust deposition, coating degradation, and temporal changes in surface emissivity or albedo can affect the module temperature $T_c$, and should be included for lifetime or seasonal studies.

\subsection*{Compact formula considering shadow}
For practical steady-state estimates, we can use equation~\eqref{eq:Tc_summary} with the understanding that $\eta$ may depend on $T_c$ and $q_{\text{cond}}=(T_c-T_{\text{mount}})/R_{\text{th}}$ if conduction is present. For high precision replace the broadband coefficients by spectral integrals and solve the spectral balance numerically.
However, equation~\eqref{eq:Tc_summary} assumes that the regolith seen by the module is fully illuminated. 
In reality, part of the regolith is shaded by the panel itself, so only a fraction of the  viewed surface contributes to reflected solar irradiance. To capture this effect, we introduce a shadow correction factor $S$ defined as
\begin{equation}
S = 1 - f_{\text{shadow}} \;\approx\; \frac{2H}{W\cos\beta + 2H}
\end{equation}
where $W$ is the panel width (projected onto the ground), $H$ the mounting height,  and $\beta$ the tilt angle. Equivalently, in terms of the aspect ratio $\Lambda = H/W$,
\begin{equation}
S \;\approx\; \frac{2\Lambda}{\cos\beta + 2\Lambda}
\end{equation}
This correction multiplies only the albedo terms, since shaded regolith still emits thermal infrared radiation according to its own temperature (we assume that shaded areas have the same temperature, which may be an approximation.

The steady-state balance taking shadow into account therefore becomes
\begin{equation}
\boxed{
T_c = \left[ 
\frac{
(\alpha_f - \eta) G_{\text{POA}}
+ S\left(\alpha_f \rho G F_{r,f} + \alpha_b \rho G F_{r,b}\right)
+ \varepsilon_{\text{reg}}\left(\varepsilon_f \sigma F_{g,f} T_{\text{reg}}^{4} 
+ \varepsilon_b \sigma F_{g,b} T_{\text{reg}}^{4}\right)
- q_{\text{cond}}
}{
(\varepsilon_f+\varepsilon_b)\sigma
}
\right]^{1/4}
}
\end{equation}

\section{Generalization Beyond the Moon}

The derived thermal model extends naturally to photovoltaic modules on any airless planetary surface or in deep space. The fundamental energy balance remains valid the same and only the environmental parameters change.

For any planet or asteroid without atmosphere, equation~\eqref{eq:Tc_summary} applies directly with updated parameters:

\begin{itemize}
  \item \textbf{Solar irradiance:} Scales with heliocentric distance as $G_{\text{planet}} = G_{\text{Earth}} \times (r_{\text{Earth}}/r_{\text{planet}})^2$, where $r$ is the orbital radius.
  \item \textbf{Surface albedo:} Material-dependent.
  \item \textbf{Surface temperature:} Varies with solar heating, thermal properties, and diurnal cycle.
\end{itemize}

The view factors $F_{r,i}$ and $F_{g,i}$ depend on mounting geometry relative to the local surface, while the solar and thermal terms scale according to the planetary environment.

In free space (no nearby planetary surface), several terms in equation~\eqref{eq:Tc_summary} vanish:
\begin{align}
\alpha_f\rho G F_{r,f} + \alpha_b\rho G F_{r,b} &= 0 \quad \text{(no reflected solar)}, \\
\varepsilon_f\sigma F_{g,f} T_{\text{reg}}^{4} + \varepsilon_b\sigma F_{g,b} T_{\text{reg}}^{4} &= 0 \quad \text{(no planetary thermal emission)}, \\
q_{\text{cond}} &= 0 \quad \text{(typically free-floating)}.
\end{align}

The thermal balance simplifies dramatically to the classic deep space result:
\begin{equation}
\boxed{ T_c = \left[ \frac{(\alpha_f - \eta) G_{\text{POA}}}{(\varepsilon_f+\varepsilon_b)\sigma} \right]^{1/4} }
\end{equation}

This represents pure solar heating balanced by radiative cooling to the cosmic background ($\sim$4~K, negligible).

\subsection{Universal applicability}

The model's physical foundation, energy conservation with radiative heat transfer, electrical conversion, and conduction, applies throughout the solar system. Whether deployed on the Moon, an asteroid mining operation, or a deep space solar power satellite, the same thermal balance governs PV module temperature. Only the environmental source terms (solar irradiance, surface albedo, surface temperature) require updating for the specific scenario. This universality makes the derived formula a valuable tool for thermal design of space-based photovoltaic systems across diverse mission architectures and destinations.

\section{Conclusion}
We have derived a general thermodynamic model to estimate the steady-state temperature of photovoltaic modules operating on the Moon. Starting from a spectral energy balance and simplifying to a practical gray-body formulation, the model captures the essential radiative and conductive processes which determine a PV module temperature in the absence of an atmosphere. The resulting compact expression incorporates measurable parameters such as plane-of-array irradiance, optical absorptance and emissivity, view factors to regolith and space, module efficiency, and conduction to the mounting system. This approach provides a clear method for predicting lunar PV operating temperatures but it can also be extended to transient conditions, spectral accuracy, and applications on other airless bodies. While we used a simple approach of an isothermal single-node, the model offers a physically justified and practical tool for system design and performance assessment of lunar photovoltaic power systems.

\bibliographystyle{unsrt}
\bibliography{references} 

\end{document}